\documentclass[prd,11pt]{revtex4-1}
\usepackage[mathcal]{euscript}
\usepackage{latexsym}
\usepackage{amsmath}
\usepackage{hyperref}
\usepackage{graphicx} 
\usepackage{verbatim} 
\usepackage{color}      
\usepackage{subfigure}  
\usepackage{hyperref}   
\usepackage{bm}
\usepackage{graphicx}
\usepackage{amssymb}
\usepackage{bbm}
\usepackage{float}
\usepackage{slashed}
\usepackage{amsfonts}
\usepackage{euscript}
\usepackage{mathrsfs} 
\usepackage{bbding}
\usepackage{pifont}
\usepackage{cancel}
\newcommand{\euler}{e}
\usepackage[utf8]{inputenc}

\makeindex

\begin{document}

\title{Black holes turn white fast, otherwise stay black: no half measures}

\author{Carlos Barcel\'o}
\email{carlos@iaa.es}
\affiliation{Instituto de Astrof\'{\i}sica de Andaluc\'{\i}a (IAA-CSIC), Glorieta de la Astronom\'{\i}a, 18008 Granada, Spain}
\author{Ra\'ul Carballo-Rubio}
\email{raulc@iaa.es}
\affiliation{Instituto de Astrof\'{\i}sica de Andaluc\'{\i}a (IAA-CSIC), Glorieta de la Astronom\'{\i}a, 18008 Granada, Spain}
\affiliation{Departamento de Geometr\'{\i}a y Topolog\'{\i}a, Facultad de Ciencias, Universidad de Granada, Campus Fuentenueva, 18071 Granada, Spain}
\author{Luis J. Garay}
\email{luisj.garay@ucm.es}
\affiliation{Departamento de F\'{\i}sica Te\'orica II, Universidad Complutense de Madrid, 28040 Madrid, Spain}
\affiliation{Instituto de Estructura de la Materia (IEM-CSIC), Serrano 121, 28006 Madrid, Spain}

\begin{abstract}{
Recently, various authors have proposed that the dominant ultraviolet effect in the gravitational collapse of massive stars to black holes is the transition between a black-hole geometry and a white-hole geometry, though their proposals are radically different in terms of their physical interpretation and characteristic time scales \cite{Barcelo2014e,Haggard2014}. Several decades ago, it was shown by Eardley that white holes are highly unstable to the accretion of small amounts of matter, being rapidly turned into black holes \cite{Eardley1974}. Studying the crossing of null shells on geometries describing the black-hole to white-hole transition, we obtain the conditions for the instability to develop in terms of the parameters of these geometries. We conclude that transitions with long characteristic time scales are pathologically unstable: occasional perturbations away from the perfect vacuum around these compact objects, even if being imperceptibly small, suffocate the white-hole explosion. On the other hand, geometries with short characteristic time scales are shown to be robust against perturbations, so that the corresponding processes could take place in real astrophysical scenarios. This motivates a conjecture about the transition amplitudes of different decay channels for black holes in a suitable ultraviolet completion of general relativity.
}
\end{abstract}
\maketitle
\flushbottom
\tableofcontents

\section{Introduction}

The construction of an ultraviolet completion of general relativity is expected to be essential for our understanding of the physics of black holes. There is a large amount of literature considering the possible ultraviolet effects in the gravitational collapse of massive stars to black holes; see \cite{Barcelo2015u} for a brief review and references therein. A possibility that has been recently considered is that black holes are converted into white holes due to ultraviolet effects, letting the matter inside them to come back to the same asymptotic region from which it collapsed. Two great advances in understanding this transition have been lately reported. 

On the one hand, {H{\'a}j{\'{\i}}cek} and collaborators \cite{Hajicek2001,Hajicek2003,Ambrus2005} considered the quantization of self-gravitating null shells, which permitted them to study the quantum-mechanical modifications of the gravitational collapse of these shells. In the context of this exact model, they have shown that a collapsing shell will bounce near $r=0$ and expand after that. In principle, this expansion lasts forever, so that the black-hole horizon formed in the gravitational collapse has to be somehow turned into a white-hole horizon, though the entire geometry describing this process cannot be easily figured out within this model. Additionally, they discussed the typical time for this bounce to occur \cite{Ambrus2005,Ambrus2004}. Their robust conclusion is that, despite some subtleties that appear in this evaluation, these times are \emph{too short} to make this phenomenon compatible with the semiclassical picture of black holes, with long-lived trapping horizons. This was considered by these authors to be an undesirable feature.

The transition between a black-hole geometry and a white-hole geometry was independently motivated in a totally different framework in the work of Barcel\'o, Garay and Jannes \cite{Barcelo2011}. More recently, a set of geometries describing this transition for a pressureless matter content was constructed, in the short essay \cite{Barcelo2014e} and its companion article \cite{Barcelo2015}. These geometries can be understood as the bounce of the matter distribution when Planckian curvatures are reached, and the corresponding propagation of a non-perturbative shock wave that goes along the distribution of matter and modifies the near-horizon Schwarzschild geometry, turning the black-hole horizon into a white-hole horizon \cite{Barcelo2015u}. Remarkably, the time scale associated with this proposal is \emph{the same} as the one obtained by Ambrus and {H{\'a}j{\'{\i}}cek} in \cite{Ambrus2005}. That the same characteristic time scale can be obtained by following rather different considerations is worth noticing. However, while in the work of {H{\'a}j{\'{\i}}cek} and collaborators this was considered to be an objectionable result, in this approach these short times are embraced as a desirable feature. The interest of this radical proposal lies in its inherent implications for our current conception of black holes, as well as the possible phenomenological implications that would follow from its short characteristic time scale \cite{Barcelo2015u,Barcelo2015}. In particular, the black-hole to white-hole transition would correspond to a transient that should lead, after dissipation is included, to stable (or metastable) objects that, in spite of displaying similar gravitational properties to black holes, could be inherently different in nature \cite{Visser2009}.

Similar geometric constructions (in the particular case of matter being described by null shells) have been explored by Haggard and Rovelli \cite{Haggard2014}, but with an essential difference. Their justification of the process is of different nature, which has a huge impact on the corresponding time scale: if the modifications of the near-horizon geometry are assumed to come from the piling-up of quantum effects originated outside but close to the horizon, much longer time scales are needed in order to substantially modify the classical geometry. Based on these indirect arguments, these authors discuss the construction of geometries describing a black-hole to white-hole transition, the characteristic time scale of which is much longer. This would permit to reconcile the black-hole to white-hole transition with the standard semiclassical picture of evaporating black holes. In consequence, this hypothetical process is quite conservative, as it essentially preserves all the relevant features of the latter picture.

Given these disparate developments, it is tempting to ask whether or not there exist theoretical arguments of different nature that could help to select which one of these processes, if any, can be realized in nature. In this paper we consider the role that  the well-known instabilities of white holes, first discussed by Eardley in \cite{Eardley1974}, play in answering this question.

\section{Eardley's instability}

From a qualitative perspective, one may argue that the reason for the instability of white holes to the accretion of matter is the following. White holes are compact gravitational objects that attract matter in the same way as their black cousins. The accumulation of the gravitationally attracted matter around a white hole provokes the bending of the light cones in its surroundings. It is not unreasonable to imagine that this bending would be similar to that occurring near a black hole. Then, matter going out from the white hole can eventually be trapped by the gravitational potential of the accreting matter, thus inhibiting the white-hole explosion and forming instead a black hole. In this section we will describe the precise mathematics behind this qualitative picture.

The original Eardley's argument \cite{Eardley1974} is formulated on the Kruskal manifold, the maximal extension of the Schwarzschild geometry \cite{Hawking1973}. His pioneering work has been followed up by a number of authors, adding new perspectives on his result, but leaving the main conclusion unchanged \cite{Barrabes1993,Ori1994,FrolovNovikov,Blau1989,Blau1989b}. Here we will also use the Kruskal manifold as the starting point though, as we will see, only some of its local properties are really essential to the discussion. Kruskal-Szekeres null coordinates permit to describe in a simple way the instability. These are defined in term of the Schwarzschild coordinates $(t,r)$ for $r>r_{\rm s}$, with $r_{\rm s}:=2GM/c^2$ the Schwarzschild radius, as
\begin{equation}
U:=-\left(\frac{r}{r_{\rm s}}-1\right)^{1/2}\exp{([r-ct]/2r_{\rm s})},\qquad V:=\left(\frac{r}{r_{\rm s}}-1\right)^{1/2}\exp{([r+ct]/2r_{\rm s})}.\label{eq:uvdef}
\end{equation}
They verify the useful relation
\begin{equation}
UV=\left(1-\frac{r}{r_{\rm s}}\right)\exp(r/r_{\rm s}).\label{eq:uvrel}
\end{equation}
There exists a different definition of these null coordinates for $r<r_{\rm s}$, but for our purposes it will not be necessary to consider explicitly that region of the Kruskal manifold. The domain of definition of these coordinates will therefore be given by $V \in (0,+\infty)$ and $U \in (-\infty,0)$. As usual, the hypersurface $V=0$ would correspond to the white-hole horizon, and $U=0$ to the black-hole horizon.

\begin{figure}[h]
	\begin{center}
		\includegraphics[width=15cm]{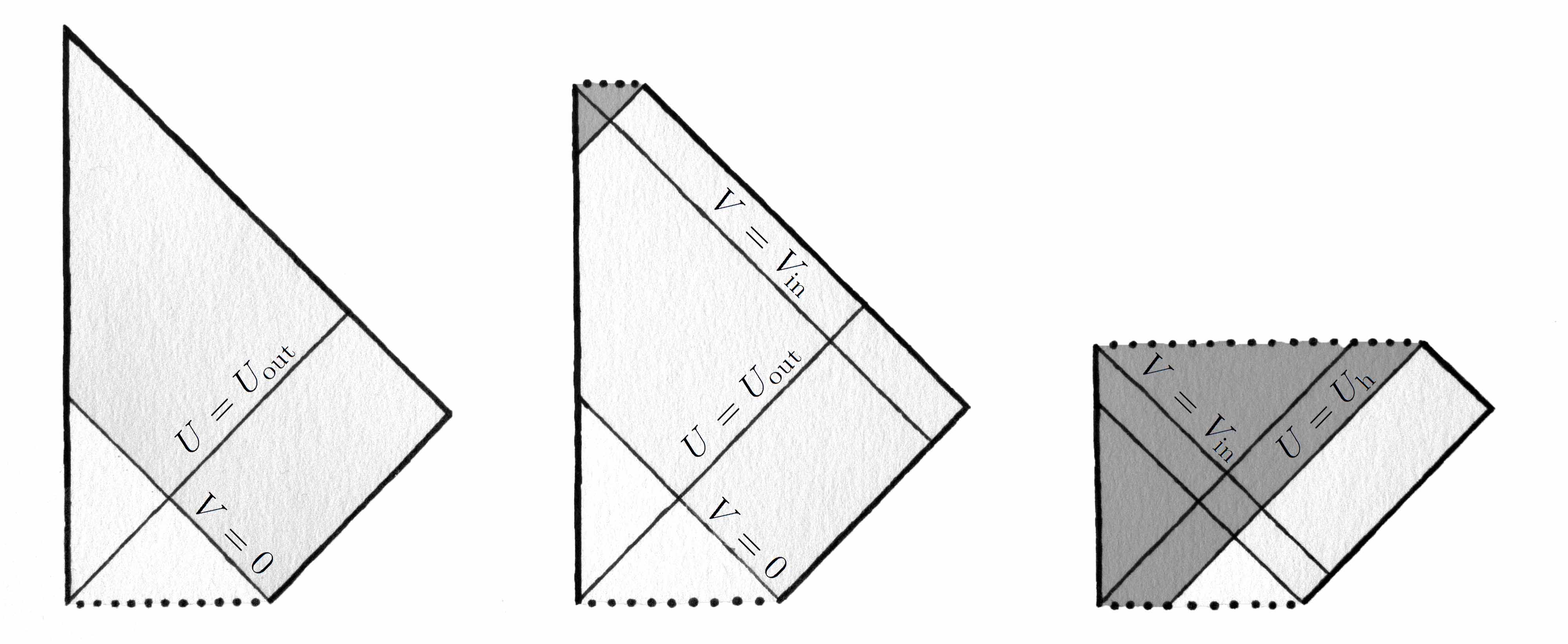}	
		\caption{Diagrammatic portrayal of Eardley's instability. The diagram on the left represents the Penrose diagram of a white hole exploding at $U=U_{\rm out}$; it is the exact time reversal of the gravitational collapse of a null shell to a black hole. On the center, the effect of an ingoing shell of matter at $V=V_{\rm in}$ is depicted, in the intuitive situation in which the crossing between the outgoing and ingoing shells occurs far from their combined Schwarzschild radius $r_{\rm s}'$. Then, essentially the entire amount of matter $M$ reaches null infinity after the explosion, and a small black hole with mass $\Delta M$ is formed. On the right, the behavior of the system when the crossing between these shells occurs below $r_{\rm s}'$ is illustrated. In these situations a black-hole horizon is formed at $U=U_{\rm h}$, the explosion of the white hole is inhibited and the outgoing shell cannot escape from $r_{\rm s}'$.\label{fig:1}}
	\end{center}
\end{figure}

The actual mathematical result that is behind this instability is the following \cite{Blau1989b}. To simplify the discussion we will use null shells to describe matter, a simplification that does not change the relevant conclusions \cite{Nunez1993}, and consider the geometry of an exploding white hole. We should think about this geometry as the exact time-reversal of the gravitational collapse of a null shell forming a black hole (see Fig. \ref{fig:1}). The two parameters of the geometry are the mass $M$ of the exploding white hole and the value $U=U_{\rm out}$ of the null shell with mass $M$ representing the outgoing matter. Now we will add to this picture an ingoing null shell, located at $V=V_{\rm in}$ and with mass $\Delta M$, representing the accretion of matter (in particular, this implies that for the region $V\geq V_{\rm in}$ one shall use a different set of null coordinates suitable glued at $V=V_{\rm in}$, but the following arguments are independent of these features).

Let us now consider the crossing between the outgoing shell and the ingoing shell. With the help of the Dray-'t Hooft-Redmount (DTR) relation for crossing null shells \cite{Dray1985,Redmount1985}, that follows from the continuity of the metric at the crossing point, one can show that a black-hole horizon is formed if the crossing between these two shells occurs below their combined Schwarzschild radius:
\begin{equation}
r_{\rm s}':=\frac{2G(M+\Delta M)}{c^2}.\label{eq:rcrit}
\end{equation}
If the crossing happens for $r=r_{\rm c}>r_{\rm s}'$ the outgoing shell can reach null infinity, but if $r_{\rm c}\leq r_{\rm s}'$ it will be inevitably confined, whatever the value $\Delta M$ takes. This is the nontrivial result that explains why the bending of light cones implies the formation of an event horizon surrounding the white hole even if only \emph{small} amounts of matter are being accreted.

In more detail, direct application of the DTR relation shows that the mass in between the two shells just after the crossing is given by
\begin{equation}
\overline{M}(\Delta r):=\frac{M+\Delta M+c^2\Delta r/2G}{\Delta M+c^2\Delta r/2G}\Delta M,\label{eq:mstar}
\end{equation}
and the energy that escapes to infinity
\begin{equation}
M_{\infty}(\Delta r):= M+\Delta M-\overline{M}(\Delta r),
\end{equation}
where $\Delta r$ marks the deviation from the value $r=r_{\rm s}'$ at the crossing, namely $r_{\rm c}=r_{\rm s}'+\Delta r$. For $\Delta r\gg r_{\rm s}'$ one has $M_\infty \simeq M$, which is the intuitive result one would expect on the basis of Newtonian considerations. However, when $\Delta r\rightarrow 0$ a genuine nonlinear phenomenon appears: the value of the energy that escapes to infinity decreases continuously down to zero. In this situation both shells, outgoing and ingoing, are confined inside a black hole with Schwarzschild radius given by Eq. \eqref{eq:rcrit}. Note that the divergence on Eq. \eqref{eq:mstar} for $\Delta r\longrightarrow-2G\Delta M/c^2$ is never reached, as the crossing between shells must occur outside the white-hole horizon, implying $V_{\rm in}>0$ or, what is equivalent, $r_{\rm c}>r_{\rm s}$. This result is independent of the particular coordinates being used, as it can be rephrased in terms of purely geometric elements: the event that corresponds to the crossing of both shells, and the properties of the asymptotic regions of spacetime. That we are using Kruskal-Szekeres coordinates is just a matter of convenience.
 
Let us make explicit how this phenomenon constrains the parameter $U_{\rm out}$ of this simple geometry. Eq. \eqref{eq:uvrel} gives us the value of the null coordinate $U=U_{\rm h}$ associated with the would-be black hole horizon formed by the accretion:
\begin{equation}
U_{\rm h}=-\frac{1}{V_{\rm in}}\frac{\Delta M}{M}\exp(1+\Delta M/M).
\end{equation}
The DTR relation implies then that the explosion of the white hole can proceed if and only if it occurs before the formation of the horizon, thus imposing a limitation on the possible values of $U_{\rm out}$:
\begin{equation}
U_{\rm out}<U_{\rm h}.
\end{equation}
It is interesting to appreciate how, in terms of these coordinates, this coordinate-invariant phenomenon is naturally expressed in a quite simple way (this is, of course, partially due to the consideration of null shells as the descriptors of matter). On the other hand, when expressed in terms of the Schwarzschild coordinates, the time for the instability to develop depends on the initial condition selected for the ingoing matter, but its order of magnitude for small accreting mass, $\Delta M\ll M$, is controlled by the typical time scale $r_{\rm s}/c\sim t_{\rm P}(M/m_{\rm P})$ \cite{Blau1989} (see the discussion in Sec. \ref{sec:extending}).

This very phenomenon does not really require of a white hole, as it only depends on the local properties of spacetime around the crossing point of the ingoing and outgoing null shells. However, its consideration as an instability could depend on the specific situation at hand. For instance, to consider a specific incarnation of white holes as unstable through Eardley's mechanism, it should combine two factors: on the one hand, white holes have to be considered as objects that explode at some point in the future; on the other hand, the matter inside the initial white-hole horizon has to be there accreting for a sufficiently long time. It is the combination of a long-standing gravitational pull, and the property of the white-hole horizon of being non-traversable from outside, that permits the accumulation of matter around the white hole for arbitrarily long times. Eardley's original consideration of white holes did contain these two factors, as he considered white holes as delayed portions of the primordial Bing Bang \cite{Eardley1974}. The discussion in the next section provides a different specific realization of white holes, as the result of the gravitational collapse of massive stars, for which the significance of the instability is even clearer.

To end this section let us mention that this classical phenomenon is not the only instability that is present in white holes. For instance, the inclusion of quantum-mechanical effects leads to further instabilities \cite{FrolovNovikov,Zeldovich1974}. For our purposes, it is enough to keep in mind that the physical reasons behind these instabilities, and therefore the corresponding typical time scales, are essentially the same as in Eardley's. Indeed, all these phenomena can be ultimately traced to the exponential behavior between the Kruskal-Szekeres and Eddington-Finkelstein coordinates. These additional sources of instabilities pile up, thus making even stronger the case for the unstable character of white holes.

\section{Black-hole to white-hole transition}

\subsection{The geometric setting}

Let us briefly discuss the relevant properties of a specific set of time-symmetric geometries describing black-hole to white-hole transitions in which matter is represented in terms of null shells initially falling from null infinity. See \cite{Barcelo2014e,Barcelo2015} for a more general discussion, with matter following timelike trajectories and different initial radii for homogeneous balls of matter. The purposes of this paper make it convenient to follow the construction in \cite{Haggard2014}, which uses Kruskal-Szekeres coordinates, though other sets of coordinates could be used instead.

\begin{figure}[h]
	\begin{center}
		\includegraphics[width=7cm]{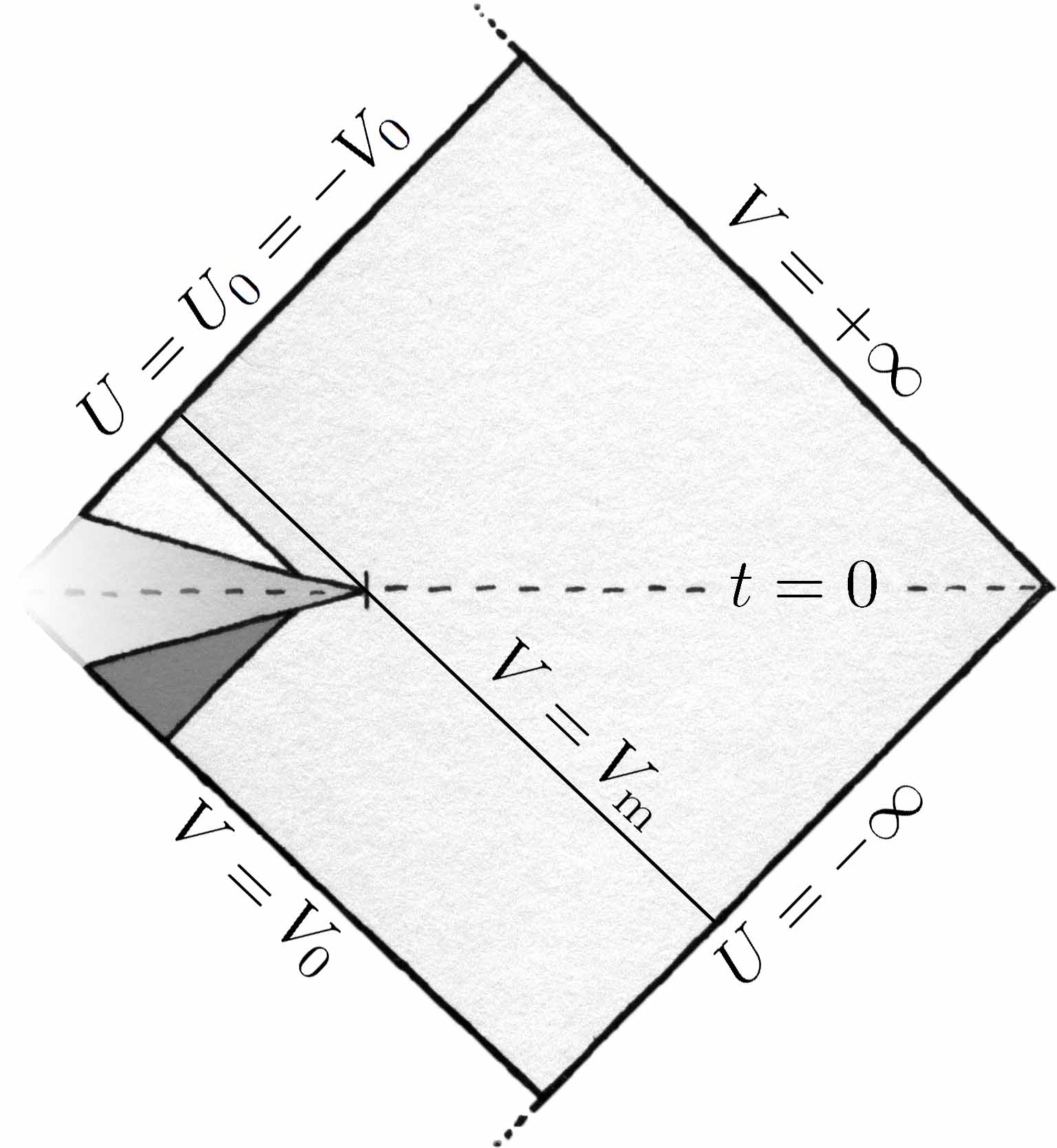}
			\caption{Part of the Penrose diagram of the spacetime describing a black-hole to white-hole transition that corresponds to the relevant portion for our analysis. The darkest gray region corresponds to the black hole interior, the lightest gray to the white hole interior. The intermediate gray region is the non-standard region interpolating between them, that extends up to $r=r_{\rm m}$ (the intersection between $V=V_{\rm m}$ and $t=0$). Part of this region fades away, representing that its explicit construction in terms of the coordinates being used is not relevant to the purposes of this paper.\label{fig:2}}
	\end{center}
\end{figure}

We will be interested in a specific region of these geometries that can be conveniently described in terms of our knowledge of the relevant patch of the Kruskal manifold. Let us consider an ingoing null shell with $V=V_0>0$ and mass $M$ as representing the collapsing matter, and another with $U=U_0:=-V_0$ and the same mass $M$ as representing its time-symmetric bounce (that would be the analogue of $U=U_{\rm out}$ in the previous section). The emphasis on the time-symmetric character of these constructions, as carried out in \cite{Barcelo2014e,Haggard2014}, is rooted in the fundamental properties of the gravitational interaction and the non-dissipative character of the bounce; the hypersurface of time reversal is conveniently labeled as $t=0$ . We are not interested here in the region under these two shells, which will be Minkowskian except in some transient region. On the other hand, we will describe the region between these two shells by means of Kruskal-Szekeres coordinates restricted to $V\in[V_0,+\infty)$ and $U\in(-\infty,U_0]$. Most of this domain will correspond to a patch of the Kruskal manifold with mass $M$. However, if the geometry were all the way down to $r=r_{\rm s}$ equal to Kruskal, the two shells will inevitably cross outside this radial position, not representing then a bounce happening inside the gravitational radius. Therefore, if we want to describe a black-hole to white-hole transition, there must exist a region in which deviations from the Schwarzschild geometry, leading to an exotic matter content, are necessary \cite{Barcelo2015,Barcelo2015u}. This region extends from the very internal region, where the classical singularity would have appeared, up to a maximum radius $r=r_{\rm m}>r_{\rm s}$ (denoted by $\Delta$ in \cite{Haggard2014}). The present knowledge of this region is quite limited, apart from the fact that it presents a \emph{physical} (i.e., there is a real physical process behind it), continuous bending of the light cones in order to continuously match the black-hole and white-hole horizons. The explicit construction of this region of spacetime is crucial for some of the properties of the transition, but not for the ones we will study here. It is interesting to note that, while using the Kruskal-Szekeres coordinates might prove powerful to understand in a simple way the part of the geometry that excludes this non-standard transition region, there are other choices of coordinates in which the \emph{entire} geometry can be constructed explicitly. We refer the reader to \cite{Barcelo2015u} in which Painlev\'e-Gullstrand coordinates are used to construct the overall geometry, including the non-standard region.

To consistently take into account that the bounce that turns the ingoing null shell into an outgoing null shell occurs inside the Schwarzschild radius, we will impose the necessary condition that $r_{\rm m}$ is greater than the crossing point between the two shells $V=V_0$ and $U=-V_0$ determined through Eq. \eqref{eq:uvrel} or, equivalently (see Fig. \ref{fig:2}),
\begin{equation}
V_0<V_{\rm m}:=\left(\frac{r_{\rm m}}{r_{\rm s}}-1\right)^{1/2}\exp(r_{\rm m}/2r_{\rm s}).\label{eq:ineq1}
\end{equation}
Here $V_{\rm m}$ marks the ingoing null shell that crosses $r=r_{\rm m}$ at $t=0$. What about the value of $r_{\rm m}$? In principle it is an unconstrained parameter that has to be fixed by independent considerations, though the natural value to consider is around $r_{\rm s}$. For instance, in \cite{Haggard2014} this quantity is identified as marking the point in which the accumulation of quantum effects outside the horizon is the largest, namely $r_{\rm m}=7r_{\rm s}/6$. On the other hand, geometries in which the initial state of the matter content corresponds to a homogeneous star with a finite radius $r=r_{\rm i}$ can be explicitly constructed by identifying the parameter $r_{\rm m}$ with $r_{\rm i}$ \cite{Barcelo2015u}. In astrophysical scenarios $r_{\rm i}$ would be always close to, and never greater than several times, $r_{\rm s}$. Therefore in all the situations we will consider, $r_{\rm m}=(1+C)r_{\rm s}$ with $C$ a dimensionless proportionality constant, which is essentially of order unity. This constant is bounded from below by $C>\mathscr{W}[V_0^2\euler^{-1}]$, where $\mathscr{W}[x]$ denotes the upper branch of the Lambert function that inverts the equation $\mathscr{W}[x]\exp(\mathscr{W}[x])=x$ and satisfies $\mathscr{W}[x]\geq -1$. None of the results of our analysis depend on the specific value that is taken for $C$, as long as it satisfies these loose constraints.

The horizon structure of the complete spacetime is shown in Fig. \ref{fig:2}. The $t<0$ of spacetime region contains a black-hole horizon that corresponds to a constant value of the null coordinate $U$, while the $t>0$ region presents a white-hole horizon with constant $V$. Note that the first trapped null ray that forms the black-hole horizon will cross the non-standard transition region and reach the future null infinity (see also Fig. \ref{fig:3}). The specific matching between the null coordinates at both sides of the non-standard transition region would need a specific ansatz for this part of spacetime (its explicit construction, in terms of Painlev\'e-Gullstrand coordinates, has been recently carried out in \cite{Barcelo2015u}). Nevertheless, the following discussion is independent of the fine details of the regularization that is taking place around the bounce, meaning that it is independent of the specific form of the metric in this region and the subsequent matching. Indeed, from all the properties of the non-standard transition region, the result will only present a mild, and as we will see irrelevant, dependence on $r_{\rm m}$ (or $V_{\rm m}$). 

For both fundamental and phenomenological purposes, the most relevant parameter characterizing a member of the above family of spacetimes is $V_0$. This parameter controls the time scale of the process for external observers. Let us consider an observer sitting at rest at $r=R\geq r_{\rm m}>r_{\rm s}$. We define the characteristic time scale for the black-hole to white-hole transition, $t_{\rm B}(R)$, as the proper time for this observer that spans between the two crossings with the ingoing and outgoing shells of matter. A standard calculation of this time using the definitions \eqref{eq:uvdef} shows that it is given by
\begin{equation}
t_{\rm B}(R)=\frac{2}{c}\sqrt{1-\frac{r_{\rm s}}{R}}\left[R+r_{\rm s}\ln\left(\frac{R-r_{\rm s}}{r_{\rm s}}\right)-2r_{\rm s}\ln(V_0)\right].\label{eq:trb0}
\end{equation}%
This function is non-negative by virtue of Eq. \eqref{eq:ineq1}. To simplify the structure of Eq. \eqref{eq:trb0} without losing the physics behind it, we will restrict ourselves to the consideration of observers that are far away from the Schwarzschild radius of the stellar structure, i.e., $r=R\gg r_{\rm s}$. The leading order in $R$ is then given by
\begin{equation}
t_{\rm B}(R)=\frac{2}{c}\left[R-2r_{\rm s}\ln(V_0)\right].\label{eq:trb}
\end{equation}%
Now we will distinguish two situations:
\begin{itemize}
\item{\emph{Short characteristic time scales:} These situations correspond to values of the parameter $V_0$ that make the time $t_{\rm B}(R)$ in Eq. \eqref{eq:trb} correspond, up to small corrections, to twice the traveling time of light from $r=R$ to $r=0$ in flat spacetime. This condition restricts the values of the parameter $V_0$ to 
\begin{equation}
V_0=V^\star_0:=\left(1-|\epsilon|\right),\qquad \epsilon\ll 1.\label{eq:bv0}
\end{equation}
These were precisely the cases considered in \cite{Barcelo2014e,Barcelo2015}. Under this condition, Eq. \eqref{eq:trb0} is coincident with the scattering times that were obtained in a quantum-mechanical model of self-gravitating null shells \cite{Ambrus2005}.
}
\item{\emph{Long characteristic time scales:} If $V_0\ll 1$, there is a significant additional delay given by
\begin{equation}
\Delta t_{\rm B}(R):=-\frac{4r_{\rm s}}{c}\ln(V_0).\label{eq:log}
\end{equation}
It is quite important to notice that the logarithmic relation between $\Delta t_{\rm B}(R)$ and $V_0$ implies that the latter has to be \emph{extremely small} in order to lead to long bounce times for asymptotic observers. This observation is essential in the following. An upper bound for the values of $V_0$ that are compatible with the time scales advocated by Haggard and Rovelli is given \cite{Haggard2014} by
\begin{equation}
V_0=V^{\rm HR}_0:=\exp(-M/2m_{\rm P})\lll 1.\label{eq:hrv0}
\end{equation}
}
\end{itemize}
Note that the only relevant time scale for the arguments in this paper is the one measured by observers that are at rest far away from the center of the gravitational potential. For other purposes, it could be interesting to evaluate the time that takes for the collapsing structure to bounce out as measured by observers attached to the star, or by observers that are located at $r=0$ always. Let us only remind that, in order to properly evaluate these quantities, it is needed to provide a specific realization of the metric of spacetime in the non-standard transition region.

\subsection{Accretion and instabilities}

Now let us describe Eardley's instability within this context. The geometries sketched above describe in the simplest possible terms the relevant aspects of the collapse and subsequent bounce of a certain amount of matter $M$ and the corresponding black-hole to white-hole transition. In the same way that we have described the collapsing matter by null shells, we will describe the accreting matter by ingoing null shells with mass $\Delta M$. An ingoing null shell of accreting matter is placed at $V=V_{\rm in}$, with $V_{\rm in}$ in principle arbitrary but always greater than $V_0$. Let us nevertheless focus for the moment on the null shells that do not traverse the non-standard transition region. Recall that the farthest radius reached by the transition zone between the black-hole and the white-hole geometries is $r=r_{\rm m}$. Matter that reaches the point $r=r_{\rm m}$ just before $t=0$ inevitably crosses the non-standard transition region, so that its evolution is not known with complete precision. On the other hand, matter that reach this point after $t=0$ does not notice anything unusual and keeps falling under the standard influence of gravity. Therefore, in order to guarantee that our arguments are model-independent, we will take for the time being the value of the null coordinates $V_{\rm in}\geq V_{\rm m}$, with $V_{\rm m}$ previously defined in Eq. \eqref{eq:ineq1}.

Now we can use Eq. \eqref{eq:uvrel} to obtain the value of the null coordinate $U$ that would correspond to the new black-hole horizon:
\begin{equation}
U_{\rm h}=-\frac{1}{V_{\rm in}}\frac{\Delta M}{M}\exp(1+\Delta M/M).\label{eq:previneq}
\end{equation}
The black-hole to white-hole transition is inhibited if and only if $U_0\geq U_{\rm h}$ which, taking into account the relation $U_0=-V_0$, is expressed in terms of $V_0$ and $V_{\rm in}$ as
\begin{equation}
V_0\leq|U_{\rm h}|=\frac{1}{V_{\rm in}}\frac{\Delta M}{M}\exp(1+\Delta M/M).\label{eq:ineq}
\end{equation}

There are different ways of extracting the physics behind these equations. Let us consider for instance an ingoing null shell, also with mass $\Delta M$, located at $V=V_{\rm in}=V_{\rm f}$ defined as the value of $V_{\rm in}$ that saturates the inequality \eqref{eq:ineq}, therefore representing the last shell that could inhibit the white-hole explosion:
\begin{equation}
V_{\rm f}:=\frac{1}{V_0}\frac{\Delta M}{M}\exp(1+\Delta M/M).\label{eq:vfdef}
\end{equation}
Now there are three disjoint situations concerning the closed interval $[V_{\rm m}, V_{\rm f}]$ in the real line:
\begin{itemize}
\item[(i)]{If this interval is empty, namely $[V_{\rm m}, V_{\rm f}]=\emptyset$, no null shell with $V_{\rm in}\geq V_{\rm m}$ could inhibit the white-hole explosion.}
\item[(ii)]{If $(V_{\rm m}, V_{\rm f})\neq\emptyset$ there is a continuum of null shells with $V_{\rm in}\in [V_{\rm m},V_{\rm f}]$ that inhibit the black-hole to white-hole transition.}
\item[(iii)]{The marginal case corresponds to $V_{\rm m}=V_{\rm f}$, in which only a specific shell located at $V_{\rm in}=V_{\rm m}=V_{\rm f}$ could forbid the transition.}
\end{itemize}
The accreting mass threshold $\Delta M$ that guarantees the condition for the marginal case (iii) to hold can be obtained by imposing $V_{\rm m}=V_{\rm f}$ and solving for Eq. \eqref{eq:vfdef}:
\begin{equation}
\Delta M=M\,\mathscr{W}\left[\euler^{-1}V_0 V_{\rm m}\right].\label{eq:wlam}
\end{equation}
Again, $\mathscr{W}[x]$ denotes the upper branch of the Lambert function. For every value of $V_0$ and $V_{\rm m}$ this equation sets a mass threshold. For masses lower than this threshold one would move to case (i), while for higher masses one would enter in case (ii).

Naturally, one can speak of an unstable behavior only in case (ii), and if the interval $[V_{\rm m},V_{\rm f}]$ is big enough in some specific sense. In order to make this possible we must consider masses $\Delta M$ that are larger than the threshold \eqref{eq:wlam}, which explains why this quantity is of singular importance to the following discussion. To unveil an intuitive measure of the size of this interval, let us rephrase the discussion in terms of a quantity with a direct physical significance, namely the proper time $\Delta T$ that an external observer sitting at $r=R\gg r_{\rm s}$ measures between these two shells, $V=V_{\rm m}$ and $V=V_{\rm f}$. Following the second of Eqs. \eqref{eq:uvdef}, this quantity $\Delta T$ is determined by means of the relation
\begin{equation}
\frac{V_{\rm f}}{V_{\rm m}}=\exp(c\Delta T/2r_{\rm s}).\label{eq:deltaTv}
\end{equation}
Now in terms of $\Delta T$, the three situations specified in the previous paragraph correspond to (i) $\Delta T<0$, (ii) $\Delta T>0$, and (iii) $\Delta T=0$.

Using Eqs. \eqref{eq:vfdef} and \eqref{eq:deltaTv} it is straightforward to obtain the following expression for $\Delta T$ in terms of the parameters of the geometry and the accreting mass $\Delta M$:
\begin{equation}
\Delta T=\frac{2r_{\rm s}}{c}\left[-\ln(V_0)-\ln(V_m)+\ln\left(\frac{\Delta M}{M}\right)+\left(1+\frac{\Delta M}{M}\right)\right].\label{eq:dtdef}
\end{equation}
Note that $\Delta T$ increases when $\Delta M$ does.

In summary, we have obtained the interval of time $\Delta T$ in Eq. \eqref{eq:dtdef}, as measured in terms of the proper time for asymptotic observers, that permits the inhibition of the white-hole explosion by the accretion of null shells of mass $\Delta M$. In terms of null coordinates, this interval corresponds to $[V_{\rm m},V_{\rm f}]$ where $V_{\rm m}$ marks the first ingoing null shell that does not cross the non-standard transition region between the black-hole and the white-hole geometry, and $V_{\rm f}$ marks the last ingoing null shell that could inhibit the transition. To guarantee that the open interval $(V_{\rm m},V_{\rm f})$ is not empty, the mass $\Delta M$ has to be greater than the threshold set by Eq. \eqref{eq:wlam}. Let us analyze in the following the effect that considering very disparate values of $V_0$ has in these expressions. The precise numbers are not essential for the discussion, but the significant result is the dramatic discrepancy between the situations with long and short characteristic time scales.

\subsection{Long vs. short characteristic time scales} 
\begin{itemize}
\item{\emph{Long characteristic time scales:} The smallness of $V=V_0^{\rm HR}$ as given by Eq. \eqref{eq:hrv0} permits, at least for $r_{\rm m}\ll r_{\rm s}|\ln(V_0^{\rm HR})|$, to use the Taylor expansion of the Lambert function in Eq. \eqref{eq:wlam} to the first nontrivial order. Taking into account that $\mathscr{W}[x]=x+\mathscr{O}(x^2)$, one has
\begin{equation}
(\Delta M)^{\rm HR}\simeq \euler^{-1} MV_0^{\rm HR} V_{\rm m}.\label{eq:mthr}
\end{equation}
This threshold is absurdly small for any reasonable value of $r_{\rm m}$ (or $V_{\rm m}$) due to the extremely tiny value of $V_0^{\rm HR}$; for instance $V_0^{\rm HR}\simeq\exp\left(-10^{38}\right)$ for stellar-mass black holes.

Now we can consider accreting masses that are several orders of magnitude higher than this threshold in order to study the properties of the finite interval in which the inhibition of the black-hole to white-hole transition could appear. For instance, we can consider the conservative lower bound
\begin{equation}
(\Delta M)^{\rm HR}\gtrsim \left(V_0^{\rm HR}\right)^\epsilon M=\exp(-\epsilon M/2m_{\rm P})M,\label{eq:lowlimt}
\end{equation}
where $\epsilon:=10^{-3}$ for instance. This threshold is still ridiculously small: it is roughly about $\epsilon M/m_{\rm P}\sim 10^{35}$ \emph{orders of magnitude} smaller than the electron mass (or any other mass scale we are used to), which clearly shows the unphysical nature of these geometries. 

Concerning the evaluation of $\Delta T$ in Eq. \eqref{eq:dtdef}, the term $\ln\left(V_0^{\rm HR}\right)$ is the dominant one by construction (at least for $r_{\rm s}\ll R\ll r_{\rm s}|\ln\left(V_0^{\rm HR}\right)|$), thus resulting
\begin{equation}
(\Delta T)^{\rm HR}\simeq-\frac{2r_{\rm s}}{c}\ln\left(V_0^{\rm HR}\right)=2\,t_{\rm P}\left(\frac{M}{m_{\rm P}}\right)^2.\label{eq:hrn}
\end{equation}
This is true for a large range of the parameter space due to the occurrence of the rest of parameters (namely $r_{\rm m}$) inside logarithms. This time interval is indeed half the characteristic time scale for the transition to develop:
\begin{equation}
(\Delta T)^{\rm HR}\simeq \frac{1}{2} t_{\rm B}(R).\label{eq:hrn1}
\end{equation}

These equations illustrate our main result concerning the black-hole to white-hole transition in long characteristic time scales: any imperceptibly small [that is, verifying Eq. \eqref{eq:lowlimt}], accidental departure from the vacuum around these geometries that takes place in the extremely long time interval given by Eq. \eqref{eq:hrn} makes it impossible for the transition to develop. This is true for a wide range of masses that encompasses both stellar-mass and primordial black holes, and independently of the unknown details of these geometries concerning the regularization of the classical singularity.
}
\item{\emph{Short characteristic time scales:} In this case the parameter $V_0=V_0^{\star}$ is restricted by Eq. \eqref{eq:bv0}. Though now the Taylor expansion of the Lambert function to the first order is not particularly useful due to its argument being larger, we can use the lower bound
\begin{equation}
(\Delta M)^\star=M\,\mathscr{W}\left[\euler^{-1}V_0^\star V_{\rm m}\right]>M\,\mathscr{W}\left[\euler^{-1}(V_0^\star)^2\right].
\end{equation}
In order to obtain this bound we have used Eq. \eqref{eq:ineq1} as well as the property of the Lambert function of being monotonically increasing for non-negative arguments. Taking into account the corresponding value of the Lambert function $\mathscr{W}[1/\euler]\simeq 0.28$, one has then
\begin{equation}
(\Delta M)^\star\gtrsim 0.28 M.\label{eq:wlamstar}
\end{equation}
The comparison between this result and Eq. \eqref{eq:mthr} shows a huge difference between these two cases. Again, we must go to larger masses than Eq. \eqref{eq:wlamstar} in order to permit a finite interval in which the white-hole explosion could be inhibited to exist but, in doing so, we rapidly reach situations with $\Delta M\simeq M$. Of course, if there is essentially the same (or even more) amount of matter going inwards that going outwards, it is not surprising that the system would tend to recollapse at first. This is a completely standard behavior that has nothing to do with an unstable behavior.
}
\end{itemize}

\subsection{Extending the analysis to the non-standard transition region \label{sec:extending}}

Our discussion could stop here, as both situations with long and short characteristic time scales have been considered on the same footing and have been shown to display very different stability properties. For ingoing null shells that do not traverse the non-standard transition region, transitions with long characteristic time scales are intrinsically unstable, while transitions with short characteristic time scales do not suffer form this pathology. For the sake of completeness we can anyway push further the discussion about the geometries with short characteristic time scales, and consider additionally the possible inhibitions of the white-hole explosion by means of null shells crossing the non-standard transition region below $r=r_{\rm m}$ at $t=0$. A precise study of these trajectories would need using a specific ansatz for this part of the geometry but, at least for our purposes, it is enough to consider the local properties of the geometry around the white-hole horizon in order to settle the stable or unstable behavior. 

For continuity reasons only, given a smooth non-standard transition region there will always be an ingoing null trajectory passing precisely through the white-hole horizon. A specifically chosen null shell placed at an ingoing null trajectory that gets sufficiently close to the white-hole horizon could therefore inhibit the explosion. Nevertheless, for short characteristic time scales this would be true for a quite limited range of initial conditions, irrespectively of whether one is considering null shells that approach the white-hole horizon from outside or from inside. The reason for this assertion lies both on the value of $V_0$ and the exponential relation between the two pairs of null coordinates, Kruskal-Szekeres $(U,V)$ and Eddington-Finkelstein $(u,v)$:
\begin{equation}
U=-\exp(-u/2r_{\rm s}),\qquad V=\exp(v/2r_{\rm s}).\label{eq:ksef}
\end{equation}
It is a well-known feature that ingoing geodesics in a white-hole geometry will pile up exponentially on the white-hole horizon (the same occurs for outgoing geodesics in a black-hole geometry). However, this accumulation is controlled by the typical scale $2r_{\rm s}$ that appears in the coordinate relations \eqref{eq:ksef}. Indeed, in the white-hole geometry this accumulation only appears when $|U|\ll 1$ or, in other words, $u\gg 2r_{\rm s}$. To explicitly show so we can use the outgoing Eddington-Finkelstein coordinates $(u,r)$ instead of the double-null Kruskal-Szekeres coordinates. Ingoing null rays still correspond to constant values of $V$ which, using Eqs. \eqref{eq:uvrel} and \eqref{eq:ksef}, is given in terms of these coordinates by
\begin{equation}
V(u,r)=\exp(u/2r_{\rm s})\left(\frac{r}{r_{\rm s}}-1\right)\exp(r/r_{\rm s}).
\end{equation}
From this relation we can directly read that ingoing null shells approach $r=r_{\rm s}$ exponentially in terms of $u$.

\begin{figure}[h]
	\begin{center}
		\includegraphics[width=15cm]{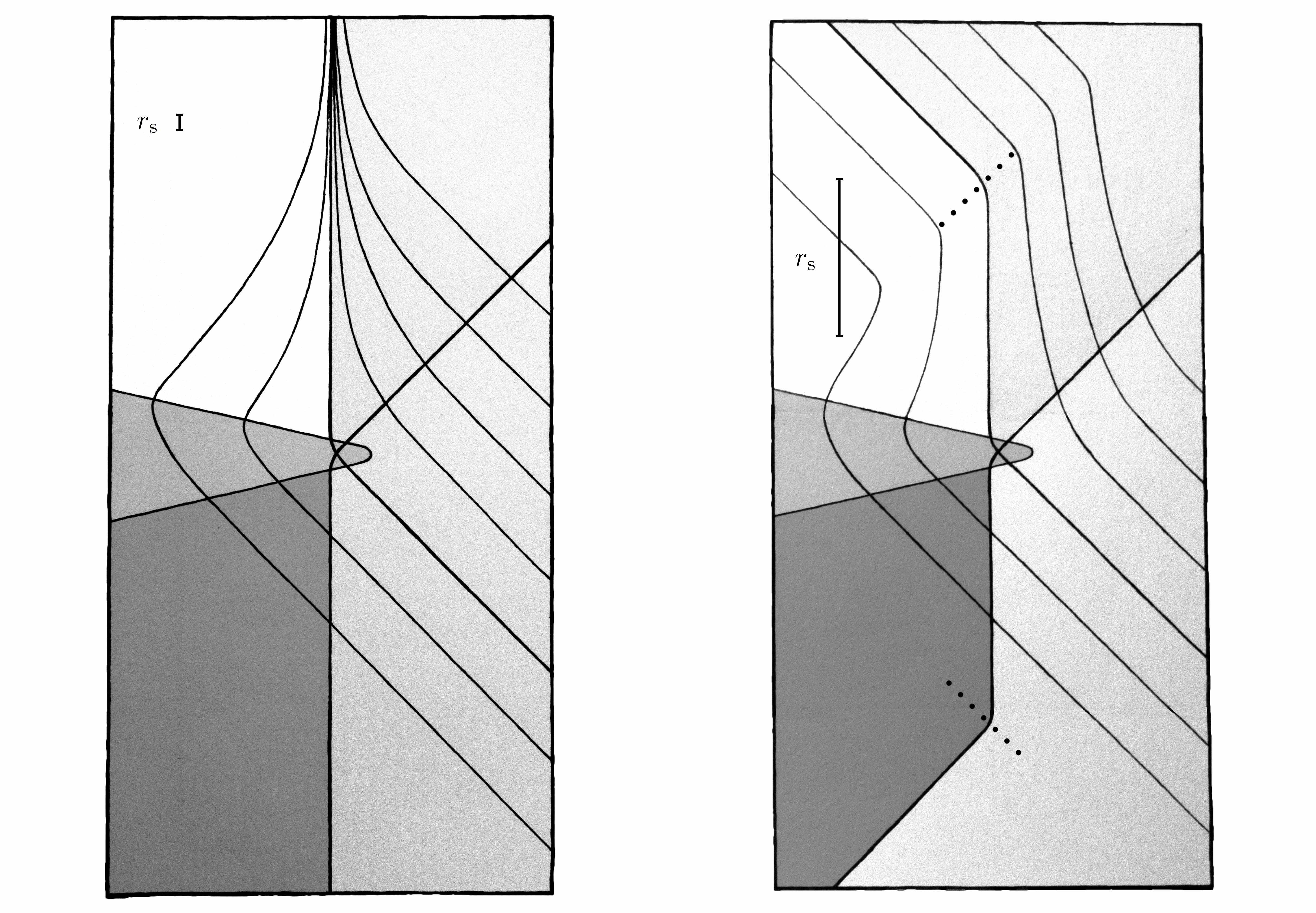}
			\caption{Illustration of the behavior of ingoing null shells on the surroundings of the white-hole horizon. On the left, the development of the exponential relation due to the extremely large value of $\Delta u^{\rm HR}$ when compared to $r_{\rm s}$ is depicted. Only part of the region of spacetime that contains both the black-hole horizon (the vertical line enclosing the darkest region) and the white-hole horizon (the vertical line enclosing the brightest region) is depicted. On the right, it is shown how much shorter durations of the white-hole horizon do not permit this exponential relation to unfold. The dotted lines represent the crossing between the ingoing and outgoing null shells describing matter (which have been omitted to highlight the relevant details) and the null rays that form the trapping horizons. The size of both diagrams is not commensurable; only the relation between the size of the white-hole horizon (as measured by the Eddington-Finkelstein coordinate $u$) and the magnitude of $r_{\rm s}$ as marked in each of the diagrams is meaningful. The actual effect is qualitative similar much more dramatic due to the extremely different values of $\ln(V_0)$ between these two cases.\label{fig:3}}
	\end{center}
\end{figure}

In the geometries describing the black-hole to white-hole transition the development of the white-hole horizon is stopped by the emission of the outgoing null shell at $U=U_0=-V_0$. This marks a stark difference between the situations with short and long characteristic time scales, as $\ln(V_0)$ turns out to represent roughly a measure of the range of initial conditions that inhibit the white-hole explosion. To realize so let us consider the interval $\Delta u$ that goes between $U=U_{\rm m}:=-V_{\rm m}$ [defined by analogy to $V_{\rm m}$ in Eq. \eqref{eq:ineq1}] and $U=-V_0$:
\begin{equation}
\Delta u=-2r_{\rm s}\ln\left(\frac{V_0}{V_{\rm m}}\right).
\end{equation}
This interval gives a natural measure of the duration of the white-hole horizon, at least when $r_{\rm m}$ is not far from $r_{\rm s}$ as we have been assuming throughout all the paper. It is now a matter of considering the different values of $V_0$ for both situations as defined in Eqs. \eqref{eq:bv0} and \eqref{eq:hrv0}; see Fig. \ref{fig:3} for a graphical representation. Let us first check that we obtain the same conclusion when analyzing the geometries with long characteristic time scales. The extremely large value of $\Delta u$ in these cases permits to fully develop the exponential accumulation of ingoing null (as well as timelike) trajectories extremely close to the long-lived white-hole horizon:
\begin{equation}
(\Delta u)^{\rm HR}=-2r_{\rm s}\ln\left(\frac{V_0^{\rm HR}}{V_{\rm m}}\right)\simeq-2r_{\rm s}\ln\left(V_0^{\rm HR}\right)\ggg 2r_{\rm s}.
\end{equation}
In this way one will find all sort of lumps of matter at arbitrarily close distances to the white-hole horizon, independently of their initial conditions, thus causing the unstable behavior. On the contrary, geometries with short characteristic time scales do not permit this accumulation or, in other words, the development of the standard exponential relation between the affine coordinates at the horizon and null infinity:
\begin{equation}
(\Delta u)^\star=-2r_{\rm s}\ln\left(\frac{V^\star_0}{V_{\rm m}}\right)=-2r_{\rm s}\ln\left(V_0^{\star}\right)+r_{\rm s}\ln\left(\frac{r_{\rm m}-r_{\rm s}}{r_{\rm s}}\right)+r_{\rm m}\lesssim r_{\rm m}\lesssim 2r_{\rm s}.
\end{equation}

In summary, it is clear from any perspective that the geometries with long characteristic time scales cannot represent a black-hole to white-hole transition in physically reasonable situations. The transition will be inevitably disrupted before being completed, and this will happen for subsequent attempts of the matter distribution to bounce out. Therefore, through all their evolution these gravitational objects will be indistinguishable from black holes.

On the other hand, for short characteristic time scales the inhibition of the white-hole explosion can be fine-tuned, but this phenomenon is far from generic and therefore we cannot speak about unstable behavior. Most importantly, even in the rare event of an initial fine-tuned inhibition of the white-hole explosion, the chances that the next bounce is also inhibited will be even more unlikely. Therefore one shall only need to wait for the next bounce (taking place in a similar time scale) that will expel the total amount of matter, both the original content and the accreted one, thus preserving the order of magnitude of the short characteristic time scale.

Let us finish this section with a few words about the generality of our results. The condition \eqref{eq:ineq} follows essentially from the application of the DTR relation, which is valid for null, spherically symmetric shells. However, it has been shown that the DTR relation can be considerably generalized to consider timelike shells of matter in spacetimes without spherical symmetry \cite{Barrabes1990}, and even massive shells \cite{Nunez1993}. Thus it seems quite safe to assume that the arguments using the DTR relation will survive these generalizations. Indeed, it is interesting to notice that essentially the same mathematics is behind the phenomenon of mass inflation, which is considered to be a robust nonlinear effect \cite{Poisson1990,Barrabes1990,Ori1991,Hamilton2010,Brown2011,Marolf2012}. As we have explained, a complementary perspective on this phenomenon is offered by the exponential relation between affine coordinates at null infinity and the white-hole horizon (which is a local property ubiquitous to white-hole horizons even in the absence of spherical symmetry), independently of the null or timelike character of the representation of the accreting matter. All these are indications that point strongly to the robust character of the instability of geometries describing the black-hole to white-hole transition with long characteristic time scales.

\section{Decay channels for black holes}

In a suitable ultraviolet completion of gravity (or, at least, of the model of spherically symmetric null shells considered here) the characteristic time scale for the black-hole to white-hole transition to develop might be calculable. A quite interesting example is given in \cite{Hajicek2001,Hajicek2003}, in which the construction of a specific quantum-mechanical model of collapsing null shells is detailed. The evaluation of the characteristic time for the transition was carried out in \cite{Ambrus2005,Ambrus2004}, supporting the scenario with short characteristic time scales; it is also interesting to read the discussion concerning the intrinsic subtleties of this evaluation. 

Even if a complete answer to this question will probably need to wait until our understanding of the ultraviolet properties of the gravitational interaction improve, there is a particular approach that might lead to a partial answer in a simple way. Within a quantum-mechanical framework, the modifications of the classical black-hole geometry may be phrased in terms of the transition between states defining different classical geometries. The initial state will be given by a black-hole geometry, and different end states will lead to different decay channels, the transition amplitudes of which can be evaluated in principle \cite{Hartle2015}. These transitions will have a characteristic timescale; for the purposes of this article, we are interested in the following processes that have been proposed:
\begin{itemize}
\item{The Hawking process \cite{Hawking1975,Hawking1976} consists in the evaporation of a black hole through the emission of particles. In geometric terms, it corresponds to the transition from a black-hole geometry to an empty (Minkowskian) geometry. The order of magnitude of the characteristic time scale for this transition can be evaluated in a semiclassical framework to be:
\begin{equation}
\mathscr{T}^{(3)}\sim t_{\rm P}\left(\frac{M}{m_{\rm P}}\right)^3.
\end{equation}
This turns out to be the less efficient channel. Given its extremely long characteristic time scale, it represents an ever present, but negligible evaporation that underlies more efficient processes.
}
\item{The process described by Haggard and Rovelli \cite{Haggard2014} represents the transition between a black-hole geometry and a white-hole geometry. These authors propose that this transition is the result of the piling-up of tiny quantum modifications, close but outside the horizon $r=r_{\rm s}$, on long enough time scales. The resulting order of magnitude for the time in which this process may occur, assuming coherent additivity (that leads to the shortest possible time), can be roughly estimated to be:
\begin{equation}
\mathscr{T}^{(2)}\sim t_{\rm P}\left(\frac{M}{m_{\rm P}}\right)^2.\label{eq:2time}
\end{equation}
While this process could be allowed in perfect vacuum, we have shown that imperceptibly small departures from the vacuum outside these geometries suffice to trigger Eardley's instability. The effect of this instability is the confinement of the matter distribution into its Schwarzschild radius, inhibiting ad infinitum the black-hole to white-hole transition.
}
\item{The works by Barcel\'o, Carballo-Rubio, Garay and Jannes \cite{Barcelo2014e,Barcelo2015} considered geometries describing a black-hole to white-hole transition, but with a completely different interpretation originally suggested in an emergent gravity context \cite{Barcelo2011}, which nevertheless stands on its own. These authors propose that the transition is caused by the propagation, in the form of a shock wave, of non-perturbative quantum effects originated near the would-be classical singularity $r=0$. We refer the reader to \cite{Barcelo2015u} for an in-depth justification of this interpretation. This realization leads to much shorter characteristic time scales for the transition to develop. The relevant time scale is essentially twice the time that light takes to cross the distance $r_{\rm s}$ in flat spacetime:
\begin{equation}
\mathscr{T}^{(1)}\sim t_{\rm P}\frac{M}{m_{\rm P}}.\label{eq:ord1}
\end{equation}
This is by far the most efficient of all these decay modes. This process circumvents Eardley's instability, so that the black-hole to white-hole transition can unfold completely.
}
\end{itemize}

These leading orders for the transition times follow a compelling pattern, being all of them of the form $\mathscr{T}^{(n)}\sim t_{\rm P}(M/m_{\rm P})^n$ for $n=1,2,3$ (note that different orders for quantum corrections would be in general classified in terms of powers of the dimensionless quantity $m_{\rm P}/M$). These leading orders include possible logarithmic corrections, that essentially preserve the order of magnitude. This leads us to the following conjecture in the particular case in which we consider an ultraviolet completion of general relativity of quantum-mechanical nature: these could be seen as different decay channels, corresponding to processes with different efficiency. Their different efficiency is reflected in their various characteristic time scales. The first possible nontrivial order for the black-hole to white-hole transition will have a characteristic time scale of the order of Eq. \eqref{eq:ord1}. Higher-order processes could lead to the illusion that much longer characteristic time scales for this transition, such as Eq. \eqref{eq:2time}, are possible in perfect vacuum. Nevertheless, we have shown that this channel is not efficient enough to drive the black-hole to white-hole transition in physically reasonable situations and is therefore forbidden in practical terms.

The most important consequence of this picture is that only in the case in which a principle or symmetry exists forbidding the faster channel, the Hawking channel could be the most relevant one. Indeed, following the standard lore in effective field theory that all which is not forbidden will inevitably happen, it seems reasonable to assume that the transitions with short characteristic time scales will generally occur, and therefore will dominate the physics of extreme gravitational collapse, unless specific conditions are imposed to forbid it. These realizations have profound implications for black hole physics \cite{Barcelo2015u}, and clearly invite to revisit and look with a different light the pioneering calculations of Ambrus and {H{\'a}j{\'{\i}}cek} \cite{Ambrus2005,Ambrus2004}, that represent the only known evaluation up to date of the parameters of the transition between a black-hole geometry and a white-hole geometry in a (simplified) model of quantum gravity.

On the other hand, while it is not strictly mandatory that any ultraviolet completion of general relativity (the nature of which does not even need to be inherently quantum-mechanical) has to permit the occurrence of black-hole to white-hole transitions in short characteristic time scales, this is indeed the only chance for this transition to take place. Otherwise, black holes will certainly keep black, with just a tiny evaporative effect due to the Hawking channel. 

\section{Conclusions}

We have considered the extension of the well-known Eardley's instability of white holes to the recently proposed geometries describing black-hole to white-hole transitions. Our main result is that the coordinate-invariant nonlinear gravitational effects encoded in the DTR relation for crossing null shells lead to specific constraints on the possible values that the parameters of these geometries could take, independently of the specific details of the regularization taking place near the bounce. In particular, transitions with long characteristic time scales, which essentially are compatible with the semiclassical picture of evaporating black holes \cite{Haggard2014}, have been shown to be pathological. In these cases, any minimal amount of matter surrounding the compact gravitational object at a particular given instant within an extremely long period of time suffocates the white-hole explosion: the amount of matter corresponding to a single typical CMB photon serves to do the job, though the critical amount of matter that is needed to trigger this instability is indeed very many (more than $10^{30}$\,!) orders of magnitude smaller. Therefore there will no explosion, nor emission of matter to the asymptotically flat region of spacetime in these cases; indeed, the bouncing distribution of matter will never be able to get outside the Schwarzschild radius. On the basis of the local geometric properties of white-hole horizons and known generalizations of the mathematical techniques being used, as well as previous work on similar phenomena, we have argued that it is quite reasonable to expect this result to be independent of the simplifications being considered (spherical symmetry and null character of the shells describing matter), ultimately being a manifestation of the nonlinear character of general relativity.

From a different conceptual perspective, one could understand the instability as the result of the fine-tuning of the parameters of the geometries describing black-hole to white-hole transitions to completely unnatural values. In particular, the pathological geometries are those for which $\ln (V_0)$ is taken to be extremely large, for instance of the order of $M/m_{\rm P}\sim 10^{38}$ for solar-mass black holes, or $M/m_{\rm P}\sim 10^{31}$ for primordial black holes. This extreme fine-tuning clashes with the rest of natural scales of the system, in particular with the characteristic time scale of Eardley's instability. When natural values are taken for the logarithm $\ln (V_0)$ this tension is alleviated: in particular, logarithms are naturally close to zero in strongly degenerate systems in which all the relevant energy scales are coincident up to small perturbations. Interestingly, this is the situation for short characteristic time scales. 

Our analysis singles out the geometries with short characteristic time scales \cite{Barcelo2015u,Barcelo2014e,Barcelo2015} as the only reasonable candidates to describe physical processes. This offers clear suggestions for ultraviolet completions of general relativity, regarding the fate of black holes and the evaluation of the characteristic time scale for the black-hole to white-hole transition, giving additional strength to the only known calculation of this quantity \cite{Ambrus2005}. Lastly, the pathological properties of the corresponding geometries compel to critically reconsider the possibility of measuring the phenomenology of hypothetical transitions with long characteristic time scales \cite{Rovelli2014,Barrau2014,Barrau2015}.


\acknowledgments
Financial support was provided by the Spanish MICINN through the projects FIS2011-30145-C03-01 and FIS2011-30145-C03-02 (with FEDER contribution), and by the Junta de Andaluc\'{\i}a through the project FQM219. R.C-R. acknowledges support from CSIC through the JAE-predoc program, co-funded by FSE, and from the Math Institute of the University of Granada (IEMath-GR).


\bibliographystyle{unsrtnat}
\bibliography{eardley_rev}	

\begin{thebibliography}{33}
\providecommand{\natexlab}[1]{#1}
\providecommand{\url}[1]{\texttt{#1}}
\expandafter\ifx\csname urlstyle\endcsname\relax
  \providecommand{\doi}[1]{doi: #1}\else
  \providecommand{\doi}{doi: \begingroup \urlstyle{rm}\Url}\fi

\bibitem[Barceló et~al.(2014)Barceló, Carballo-Rubio, and
  Garay]{Barcelo2014e}
Carlos Barceló, Raúl Carballo-Rubio, and Luis~J. Garay.
\newblock {Mutiny at the white-hole district}.
\newblock \emph{Int. J. Mod. Phys.}, D23\penalty0 (12):\penalty0 1442022, 2014.
\newblock \doi{10.1142/S021827181442022X}.

\bibitem[Haggard and Rovelli(2015)]{Haggard2014}
Hal~M. Haggard and Carlo Rovelli.
\newblock {Quantum-gravity effects outside the horizon spark black to white
  hole tunneling}.
\newblock \emph{Phys. Rev.}, D92\penalty0 (10):\penalty0 104020, 2015.
\newblock \doi{10.1103/PhysRevD.92.104020}.

\bibitem[Eardley(1974)]{Eardley1974}
Douglas~M. Eardley.
\newblock {Death of White Holes in the Early Universe}.
\newblock \emph{Phys. Rev. Lett.}, 33:\penalty0 442--444, 1974.
\newblock \doi{10.1103/PhysRevLett.33.442}.

\bibitem[Barceló et~al.(2015)Barceló, Carballo-Rubio, and
  Garay]{Barcelo2015u}
Carlos Barceló, Raúl Carballo-Rubio, and Luis~J. Garay.
\newblock {Where does the physics of extreme gravitational collapse reside?}
\newblock 2015.
\newblock URL \url{http://arxiv.org/abs/1510.04957}.

\bibitem[Hajicek and Kiefer(2001)]{Hajicek2001}
Petr Hajicek and Claus Kiefer.
\newblock {Singularity avoidance by collapsing shells in quantum gravity}.
\newblock \emph{Int. J. Mod. Phys.}, D10:\penalty0 775--780, 2001.
\newblock \doi{10.1142/S0218271801001578}.

\bibitem[Hajicek(2003)]{Hajicek2003}
Petr Hajicek.
\newblock {Quantum theory of gravitational collapse: (Lecture notes on quantum
  conchology)}.
\newblock \emph{Lect. Notes Phys.}, 631:\penalty0 255--299, 2003.
\newblock \doi{10.1007/978-3-540-45230-0_6}.
\newblock [,255(2002)].

\bibitem[Ambrus and Hajicek(2005)]{Ambrus2005}
Marcel Ambrus and Petr Hajicek.
\newblock {Quantum superposition principle and gravitational collapse:
  Scattering times for spherical shells}.
\newblock \emph{Phys. Rev.}, D72:\penalty0 064025, 2005.
\newblock \doi{10.1103/PhysRevD.72.064025}.

\bibitem[Ambrus(2004)]{Ambrus2004}
Marcel Ambrus.
\newblock \emph{{How long does it take until a quantum system reemerges after a
  gravitational collapse?}}
\newblock PhD thesis, Bern U., 2004.
\newblock URL \url{http://www.itp.unibe.ch/index.html?lang=0&id=2&subsubid=0}.

\bibitem[Barcel{\'o} et~al.(2011)Barcel{\'o}, Garay, and Jannes]{Barcelo2011}
Carlos Barcel{\'o}, Luis~J. Garay, and Gil Jannes.
\newblock {Quantum Non-Gravity and Stellar Collapse}.
\newblock \emph{Found. Phys.}, 41:\penalty0 1532--1541, 2011.
\newblock \doi{10.1007/s10701-011-9577-9}.

\bibitem[Barcel{\'o} et~al.(2015)Barcel{\'o}, Carballo-Rubio, Garay, and
  Jannes]{Barcelo2015}
Carlos Barcel{\'o}, Ra{\'u}l Carballo-Rubio, Luis~J. Garay, and Gil Jannes.
\newblock {The lifetime problem of evaporating black holes: mutiny or
  resignation}.
\newblock \emph{Class. Quant. Grav.}, 32\penalty0 (3):\penalty0 035012, 2015.
\newblock \doi{10.1088/0264-9381/32/3/035012}.

\bibitem[Visser et~al.(2009)Visser, Barce{\'o}, Liberati, and
  Sonego]{Visser2009}
Matt Visser, Carlos Barce{\'o}, Stefano Liberati, and Sebastiano Sonego.
\newblock {Small, dark, and heavy: But is it a black hole?}
\newblock 2009.
\newblock URL \url{http://arxiv.org/abs/0902.0346}.
\newblock [PoSBHGRS,010(2008)].

\bibitem[Hawking and Ellis(2011)]{Hawking1973}
Stephen~W. Hawking and George F.~R. Ellis.
\newblock \emph{{The Large Scale Structure of Space-Time}}.
\newblock Cambridge Monographs on Mathematical Physics. Cambridge University
  Press, 2011.
\newblock ISBN 9780521200165, 9780521099066, 9780511826306, 9780521099066.

\bibitem[Barrab\`es et~al.(1993)Barrab\`es, Brady, and Poisson]{Barrabes1993}
Claude Barrab\`es, Patrick~R. Brady, and Eric Poisson.
\newblock Death of white holes.
\newblock \emph{Phys. Rev. D}, 47:\penalty0 2383--2387, 1993.
\newblock \doi{10.1103/PhysRevD.47.2383}.

\bibitem[Ori and Poisson(1994)]{Ori1994}
Amos Ori and Eric Poisson.
\newblock Death of cosmological white holes.
\newblock \emph{Phys. Rev. D}, 50:\penalty0 6150--6157, 1994.
\newblock \doi{10.1103/PhysRevD.50.6150}.

\bibitem[Frolov and Novikov(1998)]{FrolovNovikov}
Valeri~P. Frolov and {\'I}gor~D. Novikov.
\newblock \emph{Black Hole Physics: Basic Concepts and New Developments}.
\newblock Fundamental Theories of Physics. Springer Netherlands, 1998.
\newblock ISBN 9780792351450.
\newblock URL \url{https://books.google.es/books?id=n0kHI6CVWZUC}.

\bibitem[{Blau} and {Guth}(1989)]{Blau1989}
Steven~K. {Blau} and Alan~H. {Guth}.
\newblock {The stability of the white hole horizon}.
\newblock \emph{Essay written for the Gravity Research Foundation 1989 Awards
  for Essays on Gravitation}, 1989.
\newblock URL
  \url{http://gravityresearchfoundation.org/pdf/awarded/1989/blau_guth.pdf}.

\bibitem[Blau(1989)]{Blau1989b}
Steven~K. Blau.
\newblock {'t Hooft Dray Geometries and the Death of White Holes}.
\newblock \emph{Phys. Rev.}, D39:\penalty0 2901, 1989.
\newblock \doi{10.1103/PhysRevD.39.2901}.

\bibitem[Nunez et~al.(1993)Nunez, de~Oliveira, and Salim]{Nunez1993}
Dario Nunez, H.~P. de~Oliveira, and Jose Salim.
\newblock {Dynamics and collision of massive shells in curved backgrounds}.
\newblock \emph{Class. Quant. Grav.}, 10:\penalty0 1117--1126, 1993.
\newblock \doi{10.1088/0264-9381/10/6/008}.

\bibitem[Dray and 't~Hooft(1985)]{Dray1985}
Tevian Dray and Gerard 't~Hooft.
\newblock {The Effect of Spherical Shells of Matter on the Schwarzschild Black
  Hole}.
\newblock \emph{Commun. Math. Phys.}, 99:\penalty0 613--625, 1985.
\newblock \doi{10.1007/BF01215912}.

\bibitem[{Redmount}(1985)]{Redmount1985}
Ian~H. {Redmount}.
\newblock {Blue-Sheet Instability of Schwarzschild Wormholes}.
\newblock \emph{Progress of Theoretical Physics}, 73:\penalty0 1401--1426,
  1985.
\newblock \doi{10.1143/PTP.73.1401}.

\bibitem[Zeldovich et~al.(1974)Zeldovich, Novikov, and
  Starobinsky]{Zeldovich1974}
Y{\'a}kov~B. Zeldovich, {\'I}gor~D. Novikov, and Alexei~A. Starobinsky.
\newblock {Quantum effects in white holes}.
\newblock \emph{Zh. Eksp. Teor. Fiz.}, 66:\penalty0 1897--1910, 1974.
\newblock URL \url{www.jetp.ac.ru/cgi-bin/dn/e_039_06_0933.pdf}.

\bibitem[{Barrabes} et~al.(1990){Barrabes}, {Israel}, and
  {Poisson}]{Barrabes1990}
Claude {Barrabes}, Werner {Israel}, and Eric {Poisson}.
\newblock {Collision of light-like shells and mass inflation in rotating black
  holes}.
\newblock \emph{Classical and Quantum Gravity}, 7:\penalty0 L273--L278, 1990.
\newblock \doi{10.1088/0264-9381/7/12/002}.

\bibitem[Poisson and Israel(1990)]{Poisson1990}
Eric Poisson and W.~Israel.
\newblock {Internal structure of black holes}.
\newblock \emph{Phys. Rev.}, D41:\penalty0 1796--1809, 1990.
\newblock \doi{10.1103/PhysRevD.41.1796}.

\bibitem[Ori(1991)]{Ori1991}
Amos Ori.
\newblock {Inner structure of a charged black hole: An exact mass-inflation
  solution}.
\newblock \emph{Phys. Rev. Lett.}, 67:\penalty0 789--792, 1991.
\newblock \doi{10.1103/PhysRevLett.67.789}.

\bibitem[Hamilton and Avelino(2010)]{Hamilton2010}
Andrew J.~S. Hamilton and Pedro~P. Avelino.
\newblock {The Physics of the relativistic counter-streaming instability that
  drives mass inflation inside black holes}.
\newblock \emph{Phys. Rept.}, 495:\penalty0 1--32, 2010.
\newblock \doi{10.1016/j.physrep.2010.06.002}.

\bibitem[Brown et~al.(2011)Brown, Mann, and Modesto]{Brown2011}
Eric~G. Brown, Robert~B. Mann, and Leonardo Modesto.
\newblock {Mass Inflation in the Loop Black Hole}.
\newblock \emph{Phys. Rev.}, D84:\penalty0 104041, 2011.
\newblock \doi{10.1103/PhysRevD.84.104041}.

\bibitem[Marolf and Ori(2012)]{Marolf2012}
Donald Marolf and Amos Ori.
\newblock {Outgoing gravitational shock-wave at the inner horizon: The
  late-time limit of black hole interiors}.
\newblock \emph{Phys. Rev.}, D86:\penalty0 124026, 2012.
\newblock \doi{10.1103/PhysRevD.86.124026}.

\bibitem[Hartle and Hertog(2015)]{Hartle2015}
James Hartle and Thomas Hertog.
\newblock {Quantum transitions between classical histories}.
\newblock \emph{Phys. Rev.}, D92\penalty0 (6):\penalty0 063509, 2015.
\newblock \doi{10.1103/PhysRevD.92.063509}.

\bibitem[Hawking(1975)]{Hawking1975}
Stephen~W. Hawking.
\newblock {Particle Creation by Black Holes}.
\newblock \emph{Commun. Math. Phys.}, 43:\penalty0 199--220, 1975.
\newblock \doi{10.1007/BF02345020}.
\newblock [167(1975)].

\bibitem[Hawking(1976)]{Hawking1976}
Stephen~W. Hawking.
\newblock {Breakdown of Predictability in Gravitational Collapse}.
\newblock \emph{Phys. Rev.}, D14:\penalty0 2460--2473, 1976.
\newblock \doi{10.1103/PhysRevD.14.2460}.

\bibitem[Rovelli and Vidotto(2014)]{Rovelli2014}
Carlo Rovelli and Francesca Vidotto.
\newblock {Planck stars}.
\newblock \emph{Int. J. Mod. Phys.}, D23\penalty0 (12):\penalty0 1442026, 2014.
\newblock \doi{10.1142/S0218271814420267}.

\bibitem[Barrau et~al.(2014)Barrau, Rovelli, and Vidotto]{Barrau2014}
Aurélien Barrau, Carlo Rovelli, and Francesca Vidotto.
\newblock {Fast Radio Bursts and White Hole Signals}.
\newblock \emph{Phys. Rev.}, D90\penalty0 (12):\penalty0 127503, 2014.
\newblock \doi{10.1103/PhysRevD.90.127503}.

\bibitem[Barrau et~al.(2015)Barrau, Bolliet, Vidotto, and Weimer]{Barrau2015}
Aurelien Barrau, Boris Bolliet, Francesca Vidotto, and Celine Weimer.
\newblock {Phenomenology of bouncing black holes in quantum gravity: a closer
  look}.
\newblock 2015.
\newblock URL \url{http://arxiv.org/abs/1507.05424}.

\end{thebibliography}

\end{document}